\documentstyle[preprint,aps]{revtex}
\newcommand{\be}{\begin{eqnarray}}
\newcommand{\ee}{\end{eqnarray}}

\begin{document}

\draft
\title{\bf Glueballs and Instantons}

\author{\bf T.~Sch\"afer and  E.V.~Shuryak}

\address{Department of Physics, State University of New York
 at Stony Brook, Stony Brook, New York 11794, USA}
\date{\today}
\maketitle
\begin{abstract}
 Gluonic correlation functions and Bethe Salpeter amplitudes are calculated
in an instanton-based model of the QCD vacuum. We consider both the pure
gauge case and the situation for real QCD with two light quark flavors.
We show that instantons lead to a strong attractive force in the
$J^{PC}=0^{++}$ channel, which results in the scalar glueball being
much smaller than other glueballs. In the $0^{-+}$ channel the
corresponding force is repulsive, and in the $2^{++}$ case it is absent.
Due to the strong classical field of the instantons, these forces
are even stronger as compared to the ones for mesons made of light quarks.
The resulting correlators, masses, coupling constants and wave functions
are compared to results of lattice simulations.
\end{abstract}
\narrowtext

   One of the most obvious question in QCD is why the observed hadrons
are made of quarks and not of gluons. It appears  that  glueballs are
much heavier than typical quark-model hadrons, and therefore they have
large widths and/or complicated decay patterns, making them difficult
to find. But why are glueballs so heavy? What are their masses, radii
and other parameters in a purely gluonic world, and how do they change
if one includes light quarks?

   To get a reference point, consider a conventional model of glueball states
such as the bag model. The lowest fermion and (electric) gluon modes in a
spherical cavity have energies $2.04/R$ and $2.7/R$. Thus, glueballs are
expected to be heavier than quark states, but not much. Ignoring
spin-dependent forces, one expects $m_{0^{++}}\simeq m_{2^{++}} \simeq 1$ GeV
and $m_{0^{-+}}\simeq 1.3$ GeV. Including these forces (and other refinements)
\cite{CHP_83} the model predicts that the low-lying glueballs have masses
$m\simeq(1.0$-$1.8)$ GeV and very similar radii $r\simeq (0.7$-$0.9)$ fm.

   A number of ``glueball candidates" have been experimentally observed, but
none was unambiguously identified (see however \cite{W_94}). Lattice
simulations provide important qualitative insights, and (although
large-scale numerical efforts are still necessary) a few statements
appear to be firmly established \cite{CSV_94}:
(i)   The lightest glueball is the scalar, with a mass in the
      1.6-1.8 GeV range.
(ii)  The tensor glueball is significantly heavier $m_{2^{++}}
      /m_{0^{++}}\simeq 1.4$, with the pseudoscalar one heavier still
      $m_{0^{-+}}/m_{0^{++}}=1.5$-$1.8$ \cite{BSH*_93}.
(iii) The scalar has a much smaller size than other glueballs.
      This is seen both from the magnitude of finite size effects
      \cite{BK_89} and  directly from measurements of the wave
      functions \cite{ISST_83,FL_92}. The size of the scalar glueball
      (defined through the exponential decay of the wave function)
      is $r_{0^{++}}\simeq 0.2$ fm, while $r_{2^{++}}\simeq 0.8$ fm
      \cite{FL_92}.
For comparison, a similar measurement for the $\pi$ and $\rho$ mesons
gives 0.32 fm and 0.45 fm \cite{FL_92}, indicating that spin-dependent
forces between gluons are stronger than between quarks.

  Important tools that provide information about gluonic interactions are the
correlation functions of gluonic operators with the relevant quantum numbers,
such as the field strength squared ($0^{++}$), the topological charge density
($0^{-+}$), and the energy density ($2^{++}$):
\be
O_S=  (gG^a_{\mu\nu})^2, \hspace{0.5cm}
O_P= \frac 12 \epsilon_{\mu\nu\rho\sigma} g^2G^a_{\mu\nu}G^a_{\rho\sigma},
     \nonumber \\
\label{def_cur}
O_T= \frac 14 (gG^a_{\mu\nu})^2-g^2 G^a_{0\alpha}G^a_{0\alpha}\, .
\ee
In the following, we consider correlation functions $\Pi_\Gamma(x)=\langle 0|
O_\Gamma(x) O_\Gamma(0)|0\rangle$ for euclidean separation $x$.  An important
low energy theorem was proven in \cite{NSVZ_79}: the integral of the scalar
correlator is determined by the gluon condensate, $\int d^4x\,\Pi_S(x)=
\frac{128\pi^2}{b}\langle (gG)^2\rangle$, where $b$ denotes the first
coefficient of the beta function and the integral is regularized by
subtracting the perturbative contribution. This theorem indicates the
presence of rather large non-perturbative corrections
in the scalar channel. On the other hand, the operator product expansion
(OPE) predicts that the leading-order power correction $O(\langle
G_{\mu\nu}^2 \rangle/x^4)$ vanishes \cite{NSVZ_79}, while radiative
corrections of the form $\alpha_s \log(x^2)\langle G_{\mu\nu}^2\rangle/x^4$,
or higher order power corrections like  $\langle gf^{abc}G_{\mu\nu}^a
G_{\nu\rho}^b G_{\rho\mu}^c\rangle/x^2$ are very small.

    In practice, there are two approaches to QCD sum rules
for scalar glueballs. In the original work \cite{NSVZ_79}, the
low energy theorem was enforced by introducing a subtraction constant.
In this case, the subtraction constant completely dominates over
ordinary power corrections, and one finds a large scalar glueball
coupling. However, the vacuum picture advocated in this paper was a rather
homogeneous one (with large size instantons melted with other vacuum
fluctuations), so that the source of the large subtraction constant
was not clear.  In more recent works on the
subject \cite{Nar_84,BS_90}, the low energy theorem was not
enforced and instead a number of higher order corrections in
the OPE were evaluated. Although the resulting mass estimate
is similar to what was obtained earlier, the resulting glueball
correlation functions are very different. In particular, the low
energy theorem is underestimated by about an order of magnitude
and the scalar glueball coupling constant is substantially smaller.

  In this paper we study gluonic correlation functions in an
instanton based model of the QCD vacuum \cite{Shu_82}. The main
assumption underlying the model is that the gauge fields in the
QCD vacuum are dominated by the strong fields of small size instantons.
Support for this assumption is provided by the calculation of hadronic
correlators in the model \cite{SSV_93} and the analysis of ``cooled"
lattice configurations \cite{CGHN_94}, where non-classical fluctuations
are removed from the vacuum. Our main point is that small size instantons
can lead to a strong enhancement in the scalar correlation function and
give a consistent description of the low energy theorem. We believe that
the smallness of the scalar glueball provides a strong argument in favor
of the picture presented here.

   Before we discuss quantitative predictions, let us consider
qualitative features of point-to-point correlators \cite{CGHN_93,Shu_cor}.
Quark correlation functions roughly fall into three classes, depending
on how asymptotic freedom is broken at intermediate distances. The
non-perturbative corrections may either be
(i)   large and attractive (the corrections have the same sign as
      the free correlation function), as for the pseudoscalar $\pi,K,
      \sigma$ mesons;
(ii)  large and repulsive, as for the heavy scalars $\eta'$ and $\delta$;
(iii) or they can be small even at rather large distances
      $x\simeq 1$ fm, as is the case for the vector mesons $\rho,a_1,
      \omega,\phi$.
This classification is easily understood \cite{Shu_cor} since the
instanton-induced interaction between quarks found by 't Hooft \cite{tH_76}
has precisely the required spin-isospin properties.

  Instanton effects in the gluonic correlation functions can be
studied by calculating the correlator in the classical field of a
single instanton. Adding the short distance contribution from
the free gluon propagator, one finds
\be
\label{free_cor}
  \Pi_{S,P}(x)&=& (\pm)\frac{384g^4}{\pi^4 x^8} + n\rho^4\Pi_{inst}(x), \\
  \Pi_T(x)    &=& \frac{24g^4}{\pi^4 x^8},
\ee
where $g$ is the running coupling constant and $\Pi_{inst}(x)$ is the
instanton contribution
\be
\label{inst_cor}
\Pi_{inst}(x)&=& \frac{12288\pi^2\rho^{-8}}{y^6(y^2+4)^5}
   \Big\{ y^8+28y^6-94y^4-160y^2-120  \nonumber \\
  \lefteqn{  + \frac{240}{y\sqrt{y^2+4}}
       (y^6+2y^4+3y^2+2){\rm asinh}\left. (\frac y2) \right\}}
\ee
with $y=x/\rho$. The approximation used is that we ignore any interference
between quantum and classical fields. The reason is that these contributions
correspond to power corrections $O(\langle G_{\mu\nu}^2\rangle/x^4)$,
which, as mentioned above, are very small in the channels considered here.

  To first order in the instanton density, we find the three scenarios
discussed above: {\em attraction} in the scalar channel, {\em repulsion}
in the pseudoscalar and {\em no} effect in the tensor channel. The last case
is a consequence of the fact that the stress tensor in the self-dual field
of an instanton is zero. If one compares the result (4) with a similar
calculation for the pion, one finds that the instanton contribution in
the glueball channel is enhanced by a factor $S_0^2=(8\pi^2/g(\rho)^2)^2$
with respect to the result for the pion. This means that despite the fact
that the scalar glueball is so much heavier than the pion, the correlation
function at distances $x\simeq\rho$ is even larger.

 In order to make these statements more quantitative and study their
dependence on the presence of light quarks, we have calculated the
correlators for three different instanton ensembles. The simplest is
the Random Instanton Liquid Model (RILM), which assumes that instantons
and anti-instantons are distributed randomly in position and color space.
Already this simple model is very successful in the description of a
large number of hadronic correlation functions \cite{SSV_93}.
In this model the QCD vacuum is dominated by small-size ($\rho\simeq 0.3$ fm)
instanton or anti-instanton fluctuations with a total density $n\simeq 1\,
{\rm fm}^{-4}$. Including the correlation between instantons introduced
by the gluonic interaction between them, gives a more complicated ensemble
which we call the Quenched Instanton Liquid Model (QILM). Also taking into
account the fermionic determinant one arrives at the unquenched Interacting
Instanton Liquid Model (IILM). As shown in \cite{SV_94}, this ensemble
reproduces an important feature of QCD : the screening of the topological
charge and a correct description of the $\eta'$ channel.

  In our simulations we calculate the correlators as in (\ref{free_cor}),
but the classical part is now evaluated for multi-instanton configurations.
The scalar correlator has an $x$-independent contribution from the gluon
condensate, which must be subtracted. The running coupling constant $g$ is
calculated from the perturbative beta function at short distances, but
frozen at a value of $\alpha_s/\pi=0.3$.

 The resulting correlation functions are shown in fig.1. The correlators
are normalized to the free ones, so that all curves approach one at
short distances. Deviations from one at intermediate distances are very
different in different channels. Up to $x\simeq 0.25$ fm, these
deviations are consistent with the single-instanton correction
(\ref{inst_cor}), but at larger $x$ multi-instanton contributions
become important. Note that for the pseudoscalar correlator in the
interacting ensemble the correction even changes sign. This is a result
of correlations between instantons and anti-instantons that lead to the
screening of the topological charge. In the ensemble with light quarks a new
state, the $\eta'$ appears. Apart from this, one observes that the results
for the three ensembles considered are rather close, suggesting that
single-instanton effects (rather than correlations among them) are dominant.

One consistency check is provided by the low energy theorem. We find
that the integral of the scalar correlation function (integrated
up to $x=0.7$ fm) is  $97\pm 6\,{\rm GeV}^4$ for the random and
$66\pm 7\,{\rm GeV}^4$ for the interacting ensembles, to be compared
with the low energy theorem value $62\,{\rm GeV}^4$. In fig.1 we
also compare our results with predictions from QCD sum rules. The
dotted lines correspond to the glueball parameters obtained in
\cite{BS_90} (the results from \cite{Nar_84} are very similar).
We clearly observe that QCD sum rules do {\em not} predict a substantial
enhancement of the scalar correlator, the value of the sum rule
is only $13\,{\rm GeV}^4$. Unfortunately, most lattice simulations use
very non-local operators (in order to increase the ground state signal) and
their correlators cannot be compared directly to our results.
We strongly encourage direct measurements of point-to-point
correlation functions and the corresponding coupling constants,
similar to the results for quark correlators reported in
\cite{CGHN_93}.

 We have fitted the glueball masses and coupling constants using a
parametrization of the spectral function that consists of a zero-width
pole and a continuum starting at a threshold $s_0$. For $\Gamma=S,P$ the
correlator reads
\be
\label{spec_rep}
  \Pi_\Gamma(x)&=& \lambda^2_\Gamma D(m_\Gamma,x) + \frac{2g^4}{\pi^2}
     \int_{s_0} ds\, s^2 D(\sqrt{s},x),
\ee
where $D(m,x)=m/(4\pi^2x)K_1(mx)$ is the euclidean scalar propagator
and $\lambda_\Gamma = \langle 0|O_\Gamma|0^{PC}\rangle$ is the coupling of
the resonance to the gluonic current.
%At large distances all correlation
%functions are expected to decay exponentially, but since we have normalized
%them to the perturbative correlators (that is, multiplied by $x^8$), the
%curves should show a maximum at $x_{max}\simeq 6.5/m$.
Statistical fluctuations at large distances limit the accuracy of our mass
determination, but the coupling constants are determined rather well.
Fitting the parametrization (\ref{spec_rep}) to the measured
correlator for scalar gluonium in the random model we find a mass
$m_{0^{++}}=1.4\pm 0.2$ GeV with the coupling strength $\lambda_{0^{++}}
=17.2\pm 0.5\,{\rm GeV}^3$. In the quenched and the full ensemble the
correlation function is somewhat smaller at intermediate distances,
a consequence of the low energy theorem discussed above. At distances
$x>0.5$ fm there are large uncertainties due to the subtraction.
The mass and coupling constant are $m_{0^{++}}=1.25$\, GeV and
$\lambda_{0^{++}}= 15.6\,{\rm GeV}^3$ in the full theory and
$m_{0^{++}}=1.75$\, GeV and $\lambda_{0^{++}}= 16.5\,{\rm GeV}^3$
in the quenched case. These values are about twice as big as the
value obtained in the only lattice measurement of this quantity,
$\lambda_{0++}=7.8\,{\rm GeV}^3$ \cite{LLL*_93} (which, according
to the authors, is compatible with the low energy theorem for glueball
masses $m\simeq 1$ GeV on the low end of their statistical errors).

  In the pseudoscalar case the classical and one-loop contributions have
opposite signs. At distances where they tend to cancel each other our
approximation (which neglects the interference between the two) becomes
questionable. However, the rapid downturn directly translates into the
position of the perturbative threshold, for which we find $\sqrt{s_0}\simeq
3.0$ GeV in the random model and $\sqrt{s_0}\simeq 2.4$ GeV in the quenched
theory. We see no clear evidence for a pseudoscalar glueball state below the
continuum threshold. In the unquenched theory we observe the $\eta'$ signal
with $m_{\eta'} \lesssim 800$ MeV and $\lambda_{\eta'}\simeq 7.0\,
{\rm GeV}^3$. Using the anomaly equation, this corresponds to $f_{\eta'}= 200$
MeV.

   The tensor channel has no large non-perturbative corrections, because
isolated instantons and anti-instantons have a vanishing energy momentum
tensor. Thus the classical contribution is entirely due to the interaction
between instantons. One may therefore question the importance of instantons
in this channel. On the other hand, the OPE also does not predict any power
corrections at $O(G^2)$ and $O(G^3)$, and in a self dual background field
all power corrections vanish \cite{NSVZ_79}: the smallness of the
non-perturbative corrections may therefore survive. We find a small
classical contribution to the tensor correlator (see fig.1), but have
not made an attempt to determine the corresponding mass value. The data
can be used to put an upper limit $\lambda_T \lesssim 0.6\,{\rm GeV}^3$
on the tensor coupling.

   Since the scalar correlator is so much bigger than the other ones,
one may speculate that the scalar glueball should also be much more
compact. We have checked this statement by calculating the
Bethe Salpeter amplitudes (or `wave functions'), defined as
\be
\label{def_wave}
\psi_\Gamma(y) = \lim_{x\to\infty} \frac{1}{\Pi_\Gamma(x)}
  \langle 0|O_\Gamma(x)O^y_\Gamma(0)|0\rangle  ,
\ee
where $O_S^y(0)=g^2 G^a_{\mu\nu}(-y /2)G^a_{\mu\nu}(y/ 2)$ etc.~are
point-split operators and $y$ is orthogonal to $x$.
%On the lattice
%\cite{FL_92} a slightly different definition of the point-split source as
%well as gauge fixing to the Coulomb gauge was used.
By definition, $\psi(0)=1$.
At small separation $x$ the wave function of the scalar glueball
essentially measures the size of the instanton, $\psi_S(y)=1-(2y/3\rho)^2
+O(y^4)$.

  Our results for the random ensemble are shown in fig.2 (those for other
ensembles are very similar). The scalar wave function is indeed found
to be very compact. It is not exponential at short distances (presumably
due to the lack of short range perturbative interactions), but the
overall shape can be described by an exponential decay $\psi(y)
=\exp(-y/R)$, with a fitted radius $R=0.21$ fm. This value is in good
agreement with the lattice result $R\simeq 0.2$ fm \cite{FL_92}. The tensor
wave function is much larger in size, $R=0.61$ fm, to be compared
with the lattice result $R\simeq 0.8$ fm. Together with our earlier work
\cite{SSV_93}, in which we determined the sizes of the pion $r=0.56$ fm and
rho meson $r=0.70$ fm, this shows that the instanton model leads to
significantly larger spin-splittings for the glueball radii. For the
pseudoscalar glueball the
interaction is repulsive and we do not find a localized wave function
in our model. We therefore conjecture that lattice measurements should
find a dip in the wave function at small distances. Let us conclude
by noting that the observed hierarchy of sizes, from a very small
scalar to a large tensor and a presumably large pseudoscalar,
is of great significance for phenomenological searches, since it
may affect the branching ratios into different final states (see
\cite{W_94} and the discussion  in \cite{FL_92}).

  In summary, we have shown that instanton-induced forces between gluons
lead to strong attraction between gluons in the $0^{++}$ channel, strong
repulsion in the $0^{-+}$ channel and no short-distance effects
in the $2^{++}$ channel. We have calculated point-to-point correlation
functions using the ``instanton liquid" model (with and without quarks).
The fitted masses are compatible with lattice results. More importantly,
our large scalar coupling constants are in agreement with the low energy
theorem, and the scalar gluonium size is as small as 0.2 fm, as it was
observed in \cite{FL_92}. We have argued that measurements of gluonic
correlators and wave functions provide important insights into the
structure of the QCD vacuum, and lattice measurements of the correlators
predicted in this paper would be very desirable.

We would like to thank P.~van Baal for a  useful discussion which triggered
this paper. This work was supported in part by the US DOE grant
No.~DE-FG02-88ER40388 and the A.~v.~Humboldt foundation.

%%%%%%%%%%%%%%%%%%%%%%%%%%%%%%%%%%%%%%%%%%%%%

\begin{figure}
\begin{center}
\leavevmode
%\epsffile{glue1.ps}
\end{center}
\caption{
Scalar, pseudoscalar and tensor glueball correlation functions normalized
to the corresponding free correlators. The results in the random, quenched
and full ensembles are denoted by stars, open triangles and solid squares,
respectively. The solid lines show the parametrization described in the text,
the dashed line the dilute instanton gas approximation, and the dotted line
the QCD sum rule calculation [10]. The horizontal line in the second figure
was added to guide the eye, the vertical scale in the third figure is
$10^{-4}$. }
\end{figure}
\begin{figure}
\begin{center}
\leavevmode
%\epsffile{glue2.ps}
\end{center}
\caption{
Scalar and tensor glueball Bethe Salpeter wave functions in the random
instanton ensemble. All wavefunctions are normalized to one at the
origin. The solid and dashed lines show a parametrization of the data
used to extract the mean square radii while the dashed curve shows the
scalar wave function in the dilute instanton gas.}
\end{figure}

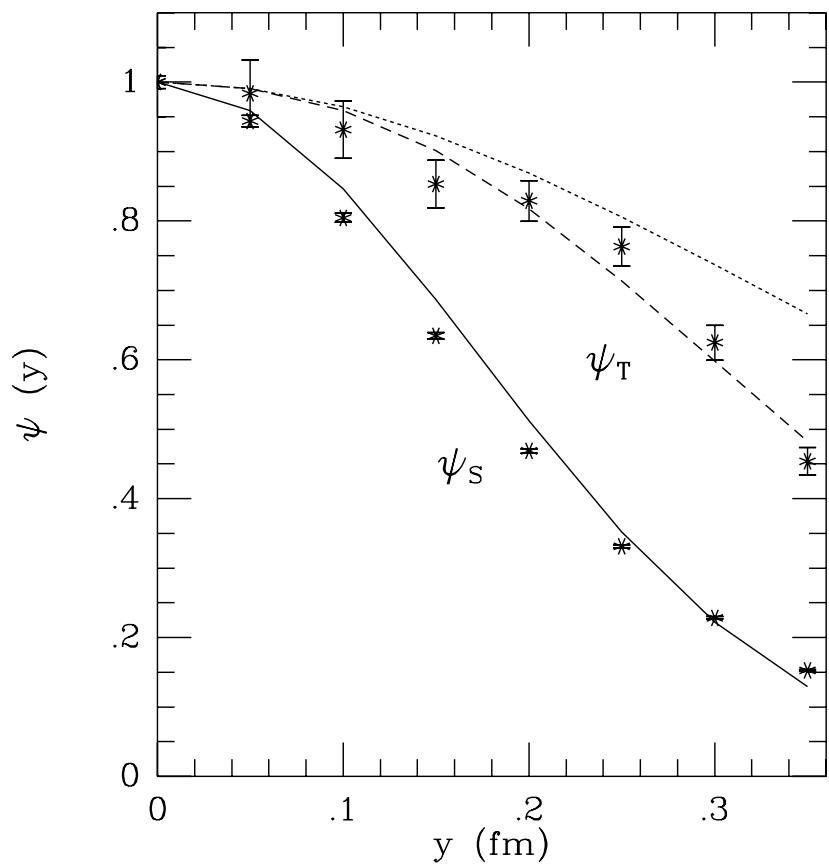
\end{document}